# Energy gaps and layer polarization of integer and fractional quantum Hall states in bilayer graphene

Yanmeng Shi[1], Yongjin Lee[1], Shi Che[1], Ziqi Pi[1], Timothy Espiritu[1], Petr Stepanov[1], Dmitry Smirnov[2], Chun Ning Lau[1*] Fan Zhang[3*]

[1] Department of Physics and Astronomy, University of California, Riverside, Riverside, CA 91765
[2] National High Magnetic Field Laboratory, Tallahassee, FL 32310
[3] Department of Physics, University of Texas at Dallas, Richardson, TX 75080

ABSTRACT

Owing to the spin, valley, and orbital symmetries, the lowest Landau level (LL) in bilayer graphene exhibits multicomponent quantum Hall ferromagnetism. Using transport spectroscopy, we investigate the energy gaps of integer and fractional quantum Hall states in bilayer graphene with controlled layer polarization. The state at filling factor $\nu$=1 has two distinct phases: a layer polarized state that has a larger energy gap and is stabilized by high electric field, and a hitherto unobserved interlayer coherent state with a smaller gap that is stabilized by large magnetic field. In contrast, the $\nu$=2/3 quantum Hall state and a feature at $\nu$=1/2 are only resolved at finite electric field and large magnetic field. These results underscore the importance of controlling layer polarization in understanding the competing symmetries in the unusual QH system of BLG.

Since the experimental isolation in 2004[1], single- and few-layer graphene has emerged as a new two-dimensional (2D) system for investigation of quantum Hall (QH) effects. The SU(4) spin-valley symmetry of these systems imparts new properties not accessible in traditional semiconductors, such as the crossover of filling factor $\nu$=0 QH state to quantum spin Hall state in tilted magnetic field[2] and the multicomponent fractional QH states[3] in single layer graphene.

For bilayer graphene (BLG), the orbital, spin and valley degrees of freedom give rise to the 8-fold degeneracy in the lowest LL [4-6], which can be broken by electronic interactions and/or single-particle perturbations, leading to QH states at intermediate integer filling factors. A particularly important parameter in exploring its symmetry-breaking QH states is the out-of-plane electric field $E_\perp$ that breaks the inversion symmetry[6-16]: as the layer and valley indices are equivalent in the lowest LL, $E_\perp$ provides an experimental "knob" for selectively breaking the layer/valley symmetry, and inducing phase transition among different ground states. For instance, the $\nu$=0 state [8,17-20] is smoothly connected to the intrinsic layer antiferromagnetic insulating state observed in ultra-clean devices[8], and evolves to a layer polarized state and a ferromagnetic state at large $E_\perp$ and $B$, respectively[21]. Even more exotic is the $\nu$=2 QH state [18,20,22-27] that hosts two distinct competing states [28]: one state has Kekulé valley order with interlayer coherence, whereas the electrons of the other state form a majority-spin quantum anomalous Hall state but minority-spin quantum valley Hall state.

* Email: lau@physics.ucr.edu; Zhang@utdallas.edu

In contrast to the well-studied QH states at $\nu$=0 and 2, the fractional and odd integer QH states in BLG are relatively less explored, due to the smaller gaps that require high mobility samples and high magnetic fields. Studies of their dependence on $B$ and $E_\perp$ have been especially limited[25,29-31], and their excitation gaps were only measured in singly-gated devices[27,29].

Here we present the transport measurements of the energy gaps for the integer QH state at $\nu$=1 and the fractional state at $\nu$=2/3 in BLG devices with controlled interlayer potential. Via transport spectroscopy on suspended dual-gated devices, we find that both states are strongly dependent on $E_\perp$. In contrast to previous studies[27,29] [30], two distinct $\nu$=1 states are resolved: phase I is resolved at $E_\perp$=0 and $B$>20T, with a relatively smaller energy gap ~0.1 meV/T, corresponding to an layer/valley coherent QH ferromagnet that is hitherto unobserved; phase II is resolved at large $E_\perp$ and relatively small $B$, with a much larger energy gap ~1.6 meV/T, corresponding to a layer/valley polarized QH ferromagnet. Our results allow us to propose specific electronic configurations for these phases. On the other hand, the $\nu$=2/3 fractional QH state is only resolved when $E_\perp$ exceeds a critical value $E_{\perp c}$, with a gap that is consistent with either a linear $B$ or a $\sqrt{B}$ dependence. A feature at $\nu$=1/2 is also only observed at finite $E_\perp$. These results highlight the importance of controlling layer polarization in understanding the interplay among quadratic band touching, spin-valley symmetry, and Coulomb interactions in the unusual QH system of BLG.

In this letter we focus on transport two BLG devices[32,33] with field effect mobilities 12,000 and 23,000 cm$^2$/Vs, and quantum mobilities ~30,000 and 40,000 cm$^{-2}$/Vs, respectively. Without symmetry breaking, the Hall conductivity of BLG is expected to be quantized at ±4, ±8, ±12… $e^2/h$, where $e$ is electron charge and $h$ Planck's constant. Fig. 1a displays the LL fan diagram $G(n, B)$ at $E_\perp$=0 for $B$=0 to 4T. LLs up to $N$=5 with properly quantized plateaus are observed (Fig. 1b) at $B$<2T, attesting to the high quality of the device. However, no symmetry-breaking QH states at -4<$\nu$<4 are observed below 6T. Interestingly, the $\nu$=6 state is resolved at $E_\perp$=0 prior to those of lower LLs, contrary to one's naive expectation. Its resolution may arise from disorder or suggest modified symmetry-breaking processes at higher LLs and warrants further studies.

We first focus on the $\nu$=1 state. Fig. 2a plots $G(n,E_\perp)$ at $B$=29T for device 1. Even at this high field, the $\nu$=1 state is not resolved at $E_\perp$=0 even at $B$=29T, but becomes fully resolved at larger $E_\perp$ (Fig. 2b). $G(E_\perp)$ displays a sharp transition – from 2.5 $e^2/h$ near $E_\perp$=0, it drops abruptly to ~ 1 $e^2/h$ when $|E_\perp|$>~10 mV/nm (Fig. 2c). In fact, at $E_\perp$=36 mV/nm, this state (and other symmetry-broken states) is resolved at $B$ as low as 3.8T, consistent with prior reports[30], thus suggesting its layer polarization in character[34,35] [36].

Strikingly, for device 2 with higher mobility, at $B$=28T, the $\nu$=1 state is resolved at both zero and finite $E_\perp$ (Fig. 2d-f). Two regions with $G$~1$e^2/h$ are visible, connected by an abrupt transition as $E_\perp$ is varied (Fig. 2e-f). Its resolution at $E_\perp$=0 is unexpected, and points to the formation of a hitherto unobserved $\nu$=1 state that is layer balanced. Our observations thus suggest the existence of two distinct phases at $v=1$: phase I is layer balanced, and appears at large $B$ and near $E_\perp=0$; phase II is layer polarized, as it is resolved at relatively small $B$, provided that $E_\perp$ exceeds certain critical value. The quantization of phase II is better than that of phase I, as seen in Fig. 2f. We note that the $v=1$ states observed in previous works almost exclusively correspond to phase II, whereas phase I has not been reported before (see Supplemental Material for discussion).

To explore these two distinct phases, we measure the scaling of their LL gaps on $B$ by using the source-drain bias $V$ as a spectroscopic tool[19,28], which has been applied to measuring the gaps of the single particle $v=4$ state[19] and the two competing correlated $v=2$ states in BLG[28]. Fig 3a plots $G(V,E_\perp)$ for device 2. The bright white/brown area near $E_\perp=0$ corresponds to the layer-balanced phase I, which abruptly gives way to the blue regions at larger $|E_\perp|$ that corresponds to the layer-polarized phase II. The line traces $G(V)$ at $E_\perp=0$ and -20 mV/nm are shown in Fig. 3b. Both display conductance valleys, yet their widths differ considerably. The LL gaps are extracted by measuring the full width half maximum (FWHM) of the valley, which is fitted to a Gaussian function. The resultant values are shown in Fig. 3c as functions of $B$. For the layer polarized phase II, the LL gap scales linearly with $B$ as $\Delta_{II}(B)\sim1.6$ meV/T, consistent with a previous work[27]. Rather unexpectedly, it appears that $\Delta_{II}$ has little or no dependence on $E_\perp$, as data at $E_\perp=-20$ and -40 mV/nm yield almost identical results. In contrast, the LL gap of phase I at $E_\perp=0$ is at least one order of magnitude smaller, though it also scales linearly with $B$ as $\Delta_I(B)\sim0.1$ meV/T.

Our experiments provide the first measurement of the LL gap for the $v=1$ state with controlled $E_\perp$. The layer balanced phase I is observed here for the first time, and is most likely a coherent linear combination of the top and bottom layers, or equivalently, K and K' valleys, since it is stable at $E_\perp=0$. Phase II is only resolved for $E_\perp>E_{\perp c}$, which is $\sim 15$ mV/nm at $B=20$T, and is evidently layer polarized. Phase II is likely the one observed in singly-gated devices[27,30] (with the possible exception of the low field data in ref. [29], see discussion in Supplemental Materials). Both phases at $v=1$ are interactions induced QH ferromagnetic states[23,24,27,37], and correspond to filling one of the two levels in an orbital doublet (LL index $N=0,1$) that has the same spin-valley index. They are energetically favored by gaining exchange energies that approximately scale as $e^2/\ell_B^2\sim B$ for *screened* Coulomb interactions. More exchange energies are gained when a LL is layer polarized, as the intra-layer exchange is generally larger than the inter-layer exchange. These two features qualitatively explain the observed linear $B$ dependence of $\Delta_I$ and $\Delta_{II}$, their relative magnitudes $\Delta_I<\Delta_{II}$, and the better quantization of phase II.

We now discuss the microscopic nature of the two phases at $v=1$. In the lowest LL, while the orbital degeneracy must be broken, the relative order and magnitude of polarizing the real spin and the layer/valley pseudospin characterize the corresponding phase. Phase I that appears at large $B$ and vanishing $E_\perp$ can be uniquely determined[36] (Fig. 3d, upper panel); the real spin polarization is maximized whereas for each occupied LL the pseudospin is layer/valley symmetric or antisymmetric. On the other hand, there are two possible candidates for phase II [30,34,36]. In the first scenario depicted as Phase II-a, the pseudospin may be maximized first because of the presence of large $E_\perp$, followed by the maximization of real spin in the last occupied $N=0$ LL. Alternatively, in the second scenario of Phase II-b, which similar to that proposed in ref. [30], the real spin may be maximized first because of the presence of large $B$, followed by the maximization of pseudospin in the last filled $N$=0 LL. Both candidates of phase II are pseudospin and spin polarized, albeit the former has a larger pseudospin polarization while the latter has a larger real spin polarization. Further experiments will be necessary to determine which candidate corresponds to the observed phase II and to explore the possible quantum phase transition between the two candidates.

The observation of the two phases, one appearing at small $E_\perp$ and large $B$ and one resolved at small $B$ and stabilized by large $E_\perp$, is reminiscent of the two competing phases of the $\nu$=2 state[28], though with one important distinction: the gap of the layer polarized $\nu$=2 state extrapolates a finite intercept at $B$=0, whereas both phases of the $\nu$=1 state appear to extrapolate vanishing gaps at $B$=0. The former feature is consistent with the fact that the layer polarized $\nu$=2 state survives to anomalously weak $B$ and adiabatically connects to the spontaneous QH state in BLG at B=0 with the same Hall conductivity[38]. In contrast, similar correspondence is absent for the $\nu$=1 state, which requires a splitting between the $N$=0 and $N$=1 LLs and thus has no counterpart in the $B$=0 limit. Therefore, both phases of the $\nu$=1 state are only resolved above their corresponding critical magnetic fields.

We now turn to the fractional QH state at $\nu$=2/3. Similar to the $\nu$=1 state, it is strongly dependent on $E_\perp$. At $E_\perp$=0, it is unresolved even at the highest attainable field $B$=31T. However, it is resolved in the presence of an interlayer potential that breaks the inversion symmetry. Fig. 4a plots the differentiated conductance $dG/dn(B,\nu)$ of device 1 at $E_\perp$=35 mV/nm. The QH plateaus appear as vertical white strips centered at various given $\nu$. Apart from the complete lifting of the 8-fold degeneracy of the lowest LL, features between $\nu$=0 and 1 are also observed – in particular, thin vertical strips at $\nu$=1/2 and 2/3 are evident. These features are more clearly visible in Fig. 4c, which displays the line traces $G(n)$ curves at $B$=21, 24, 27 and 31 T, where a small plateau at $\nu$=2/3 appears. These curves collapse into a single curve when plotted as $G(\nu)$ (Fig. 4d). This is in sharp contrast to the line trace at $E_\perp$=0 (Fig. 4c-d, dotted line), in which only $\nu$=0 and $\nu$=2 plateaus are resolved (for device 1). We note that the conductance is not perfectly quantized, presumably due to the non-zero $\sigma_{xx}$ signals (common for FQH states with small charge gaps) that are included in the two-terminal measurements.

A similar data set exhibiting the $\nu$=2/3 QH state in device 2 is shown in Fig. 4b, which plots $dG/dV_{bg}$ $(B, \nu)$ for $B$=24 to 31T. Here only the back gate is engaged; thus at finite densities, partial screening by electrons on the bottom layer leads to charge imbalance and hence finite interlayer $E_\perp$, which estimated to be ~35 – 45 mV/nm for the measurement. The observation of a clear FQH state at $\nu$=2/3 is consistent with data from device 1, namely, the $\nu$=2/3 state is only resolved at finite $E_\perp$, and in agreement with a previous work[30].

To further explore the field dependence of the $\nu$=2/3 fractional QH state, we measure $G(V, E_\perp)$ at $B$=21T and $\nu$=2/3 for device 1 (Fig. 4e). At $E_\perp$=0, $G$~2 $e^2/h$, indicating that neither $\nu$=1 nor $\nu$=2/3 states are resolved, as discussed above; as $E_\perp$ exceeds a critical value ($E_{\perp c}$ ~17 mV/nm), $G$ drops abruptly to ~0.7 $e^2/h$ (Fig. 4f). Such a dramatic transition in $G$ induced by $E_\perp$ is rather similar to that of the $\nu$=1 state, though their critical $E_\perp$ values differ. We note that a similar $E_\perp$–induced transition has been observed[30]; what sets our work apart is that, due to the higher resolution of our data, what appeared as a single transition point at $E_\perp$=0 in ref. [30] is resolved to be a broadened plateau with well-defined $E_{\perp c}$ values, hence clarifying the necessity of to support the $\nu$=2/3 state.

At $\nu=2/3$, for $E_\perp$ below (exceeds) $E_{\perp c}$, the $G(V)$ curves display zero bias conductance peaks (valleys), indicating that the fractional QH state is unresolved (resolved) (Fig. 4g). The LL gap $\Delta_{2/3}$ is extracted by measuring the FWHM of the resolved conductance valley at different $B$ and constant $E_\perp=35$ mV/nm. The resultant values, as plotted in Fig. 4h, are consistent with linear $B$ dependence, with the best-fit equation, $\Delta(B) = -2.03 + 0.16\, B[T]$ meV. However, due to the limited range in $B$, a $\sqrt{B}$ dependence cannot be definitively ruled out, where the data points may also be fitted to $\Delta(B) = -6.0 + 1.6\sqrt{B[T]}$ meV; other functional dependence may also be possible.

In the standard picture, the FQH states arise from the quench of kinetic energy by a strong magnetic field and the presence of electron-electron interactions. The negative intercepts at $B=0$ reflects the former requirement. For long-range Coulomb interactions the FQH gaps should scale with $e^2/\ell_B \sim \sqrt{B}$. The possible linear $B$ dependence is related to the strong screening that yields similar gaps scaling with $e^2/\ell_B^2 \sim B$ of the integer broken symmetry QH states in BLG. For instance, similar linear $B$ dependence is observed above for both phases of the $\nu=2$ state[28]. Future experiments in higher-mobility samples at larger fields will be necessary to confirm whether multiple phases exist at $\nu=2/3$[30] and whether the composite fermions undergo similar transitions in real spin and pseudospin polarizations to those electrons in the $N=0$ LL of the $\nu=1$ state.

To summarize, in high mobility samples we observe two distinct $\nu=1$ states and one $\nu=2/3$ state. At $\nu=1$, phase I is resolved at small $E_\perp$ and large $B$, with a LL gap of 0.1 meV/T and possible interlayer coherence; phase II is resolved at weak $B$ and large $E_\perp >\sim$-10 mV/nm, with a much larger LL gap of 1.6 meV/T and at least partial layer polarization. For the $\nu=2/3$ state, a similar dependence on $E_\perp$ is observed, though the state is only resolved for $E_\perp >\sim 20$ mV/nm, with a LL gap that rises from 1.2 to 2.8 meV as $B$ increases from 20 to 30T. Our data are consistent with prior results[27,29] and can also account for the super-linear dependence of the LL gaps observed[29] (see Supplemental Materials for a detailed discussion).

Finally, we note that our data exhibit a tantalizing feature at $\nu=1/2$, which appears as a thin white band in Fig. 4b and a small kink in the line traces; similar to the $\nu=2/3$ state, it disappears at $E_\perp=0$. Simply to compare their orbital nature, the $\nu=1/2$ state in BLG is likely similar to the $\nu=1/2$ state in conventional GaAs/AlGaAs heterostructures, which is a Fermi liquid instead of a QH state (unlike the observed $\nu=-1/2$ state in BLG which might be described by the non-Abelian Moore-Read state[31,39]). Since the two-terminal geometry of our devices convolves longitudinal and transverse signals, we are unable to conclusively determine whether a QH state is evident at $\nu=1/2$ or to relate the observed dependence on $E_\perp$ to a Fermi liquid. We note that a feature at $\nu=1/2$ state was recently observed in singly-gated devices using a transconductance fluctuation technique[40], though its nature was similarly undetermined. Further experimental studies will be necessary to explore its dependence on $E_\perp$ and to ascertain the nature of this intriguing even-denominator state.

We thank Yafis Barlas for discussion. This work is supported by DOE BES Division under grant no. ER 46940-DE-SC0010597. F.Z. is supported by UT Dallas research enhancement funds. Additional support for nanofabrication is provided by CONSEPT center at UCR.

FIG. 1 (a). *dG/dn(B, n)* between *B* = 0T and 4T. Arrows and numbers indicate filling factors. (b) *G(n)* traces at several *B*.

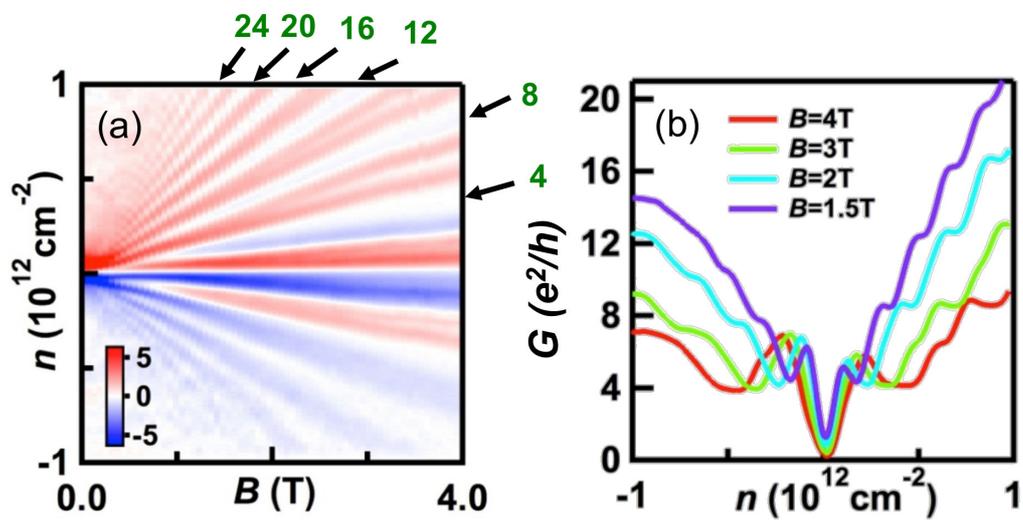

FIG. 2. (a-c). Data from device 1 at $B=29$T: $G(E_\perp,\nu)$, $G(\nu)$ line traces at zero (blue) and finite $E_\perp$ (red), and $G(E_\perp)$ line trace along $\nu=1$ (green). (d-f). Similar data set from device 2 with higher mobility at $B=28$T.

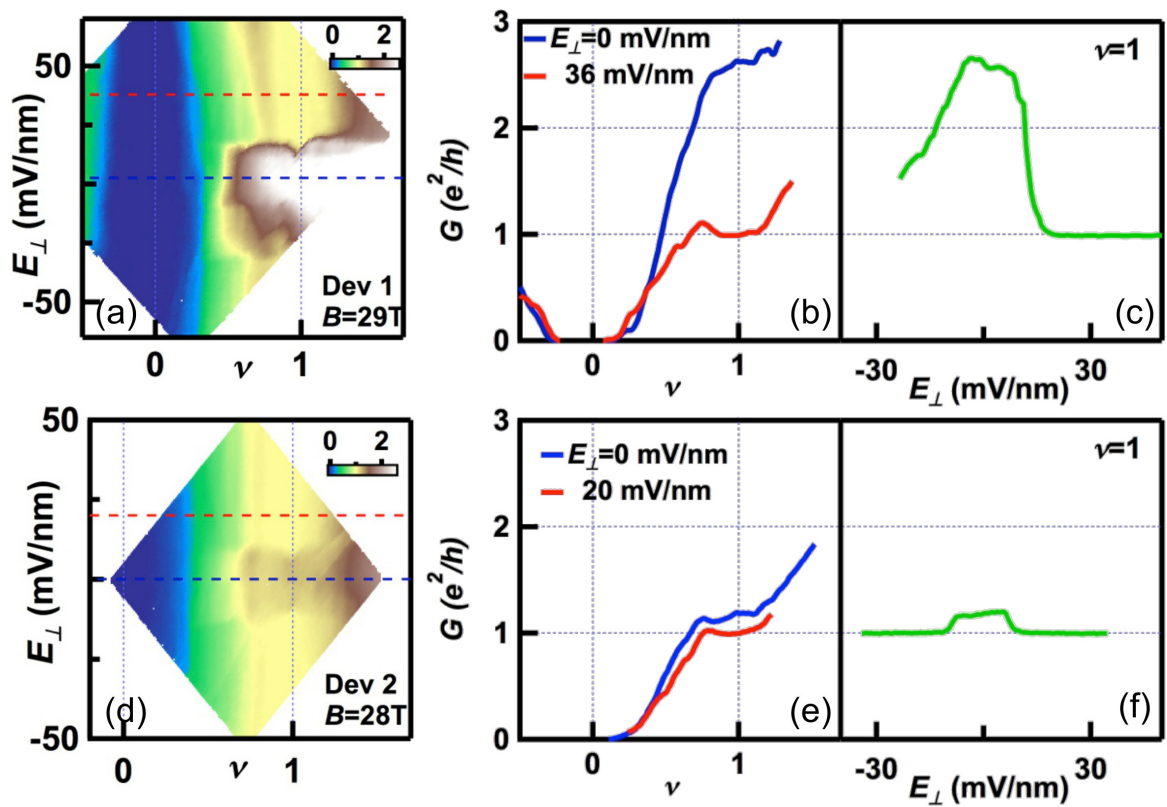

FIG. 3 (a) $G(V, E_\perp)$ of Device 2 at $B = 20T$. (b) $G(V)$ line traces at $E_\perp=0$ and $E_\perp=-20$mV/nm at 23T. (c) Measured LL gap $\Delta(B)$ at $E_\perp=0$ (blue) and $E_\perp=-20$mV/nm and $-40$mV/nm (red) respectively. (d). Schematics of electronic configurations of the different $\nu=1$ phases. T (B): top (bottom) layer; S (AS): their symmetric (anti-symmetric) combination; 0 and 1 are the orbital indices; solid (dotted) lines represent occupied (empty) levels.

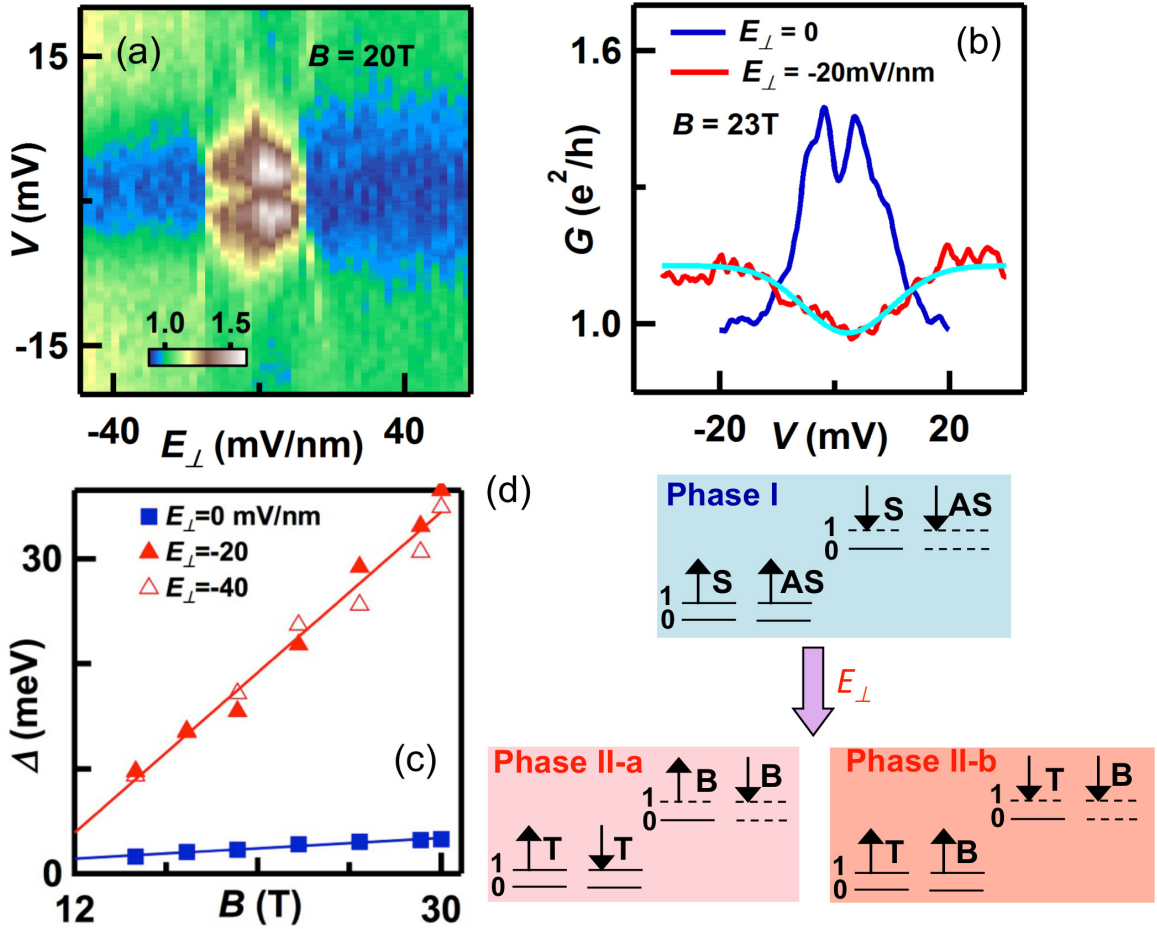

FIG. 4 (a) *G(B, ν)* of device 1 at $E_\perp$=35mV/nm. (b) *G(B, ν)* of device 2 with back gate only. (c-d) Solid lines: *G(n)* and *G(ν)* at $E_\perp$ = 35mV/nm and *B* = 21T, 24T, 27T, 31T, respectively. Dotted lines: data of device 1 at $E_\perp$=0 and *B*=29T. (e) *G(V, $E_\perp$)* and *G($E_\perp$)* at *V*=0 line trace of device 2 at *B*=21T and ν=2/3. (g) Line traces of *G(V)* at $E_\perp$=17mV/nm (blue) and $E_\perp$=25mV/nm (brown). (h) LL gap of ν=2/3 state versus *B* field. Symbols: data. Blue dot line and brown solid lines are fits using linear *B* and $\sqrt{B}$ dependences, respectively.

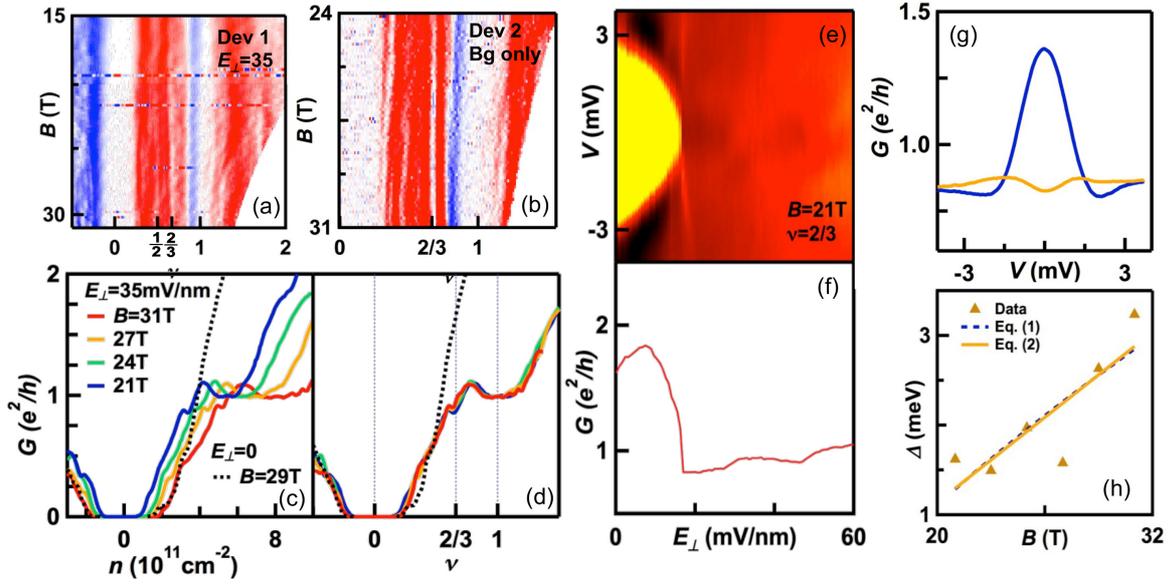